\documentclass[12pt,preprint,longabstract]{aastex}

\shorttitle{Detection of Millimeter Emission from the Circumstellar Dust
Disk Around \mbox{V1094\ Sco}}
\shortauthors{Tsukagoshi et al.}

\begin{document}


\title{Detection of Strong Millimeter Emission from the Circumstellar Dust Disk Around \mbox{V1094\ Sco}: Cold and Massive Disk around a T Tauri Star in a Quiescent Accretion Phase?}


 \author{%
   Takashi TSUKAGOSHI\altaffilmark{1}
   Masao SAITO\altaffilmark{2}
   Yoshimi KITAMURA\altaffilmark{3}
   Munetake MOMOSE\altaffilmark{4}
   Yoshito SHIMAJIRI\altaffilmark{5,6}
   Masaaki HIRAMATSU\altaffilmark{7,8}
   Norio IKEDA\altaffilmark{3}
   Kazuhisa KAMEGAI\altaffilmark{3}
   Grant WILSON\altaffilmark{9}
   Min S.~YUN\altaffilmark{9}
   Kimberly SCOTT\altaffilmark{9}
   Jay AUSTERMANN\altaffilmark{9}
   Thushara PERERA\altaffilmark{9}
   David HUGHES\altaffilmark{10}
   Itziar ARETXAGA\altaffilmark{10}
   Philip MAUSKOPF\altaffilmark{11}
   Hajime EZAWA\altaffilmark{5}
   Kotaro KOHNO\altaffilmark{1,12}
   and
   Ryohei KAWABE\altaffilmark{5}
}

 \altaffiltext{1}{Institute of Astronomy, Faculty of Science, University
 of Tokyo, Osawa 2-21-1, Mitaka, Tokyo, 181-0015, Japan: ttsuka@ioa.s.u-tokyo.ac.jp}
 \altaffiltext{2}{National Astronomical Observatory of Japan, Osawa
   2-21-1, Mitaka, Tokyo 181-8588, Japan}
 \altaffiltext{3}{Institute of Space and Astronautical Science, Japan
   Aerospace Exploration Agency, Yoshinodai 3-1-1, Sagamihara, Kanagawa
   229-8510, Japan}
 \altaffiltext{4}{Institute of Astrophysics and Planetary Sciences,
   Ibaraki University, Bunkyo 2-1-1, Mito 310-8512, Japan}
 \altaffiltext{5}{Nobeyama Radio Observatory, Nobeyama 462-2, Minamimaki,
   Minamisaku, Nagano 384-1305, Japan. Nobeyama Radio Observatory (NRO)
   is a branch of the National Astronomical Observatory of Japan (NAOJ)}
 \altaffiltext{6}{Department of Astronomy, School of Science, University
   of Tokyo, Hongo 7-3-1, Bunkyo, Tokyo 113-0033, Japan}
 \altaffiltext{7}{Academia Sinica Institute for Astronomy and
   Astrophysics, P.O. Box 23-141, Taipei 10617, Taiwan}
 \altaffiltext{8}{Institute of Astronomy, National Tsing Hua University,
   101 Section 2 Kuang Fu Road, Hsinchu 30013, Taiwan}
 \altaffiltext{9}{Department of Astronomy, University of Massachusetts,
   Amherst, MA 01003}
 \altaffiltext{10}{Instituto Nacional de Astrof\'isica, \'Optica y
   Electr\'onica, Luis Enrique Erro \#1, Tonantzintla, Puebla, M\'exico}
 \altaffiltext{11}{Department of Physics and Astronomy, Cardiff University, Cardiff CF24 3YB, Wales, UK}
 \altaffiltext{12}{Research Center for the Early Universe, School of Science, University of Tokyo, 7-3-1 Hongo, Bunkyo, Tokyo 113-0033, Japan}



\begin{abstract}
We present the discovery of a cold massive dust disk around the T Tauri star \mbox{V1094\ Sco} in the Lupus molecular cloud from the 1.1 millimeter continuum observations with AzTEC on ASTE.
A compact ($r\lesssim$320 AU) continuum emission coincides with the stellar position having a flux density of 272 mJy which is largest among T Tauri stars in Lupus.
We also present the detection of molecular gas associated with the star in the five-point observations in $^{12}$CO J=3--2 and $^{13}$CO J=3--2.
Since our $^{12}$CO and $^{13}$CO observations did not show any signature of a large-scale outflow or a massive envelope, the compact dust emission is likely to come from a disk around the star.
The observed SED of \mbox{V1094\ Sco} shows no distinct turnover from near infrared to millimeter wavelengths, which can be well described by a flattened disk for the dust component, and no clear dip feature around 10 $\micron$ suggestive of absence of an inner hole in the disk.
We fit a simple power-law disk model to the observed SED.
The estimated disk mass ranges from 0.03 to $\gtrsim$0.12 $M_\sun$, which is one or two orders of magnitude larger than the median disk mass of T Tauri stars in Taurus.
The resultant temperature is lower than that of a flared disk with the well-mixed dust in hydrostatic equilibrium, and is probably attributed to the flattened disk geometry for the dust which the central star can not illuminate efficiently.
From these results together with the fact that there is no signature of an inner hole in the SED, we suggest that the dust grains in the disk around \mbox{V1094\ Sco} sank into the  midplane with grain growth by coalescence and is in the evolutional stage just prior to or at the formation of planetesimals.
\end{abstract}



\section{Introduction}
Protoplanetary disk around pre-main sequence stars are potential sites for planet formation.
In the standard model \citep[e.g.][]{bib:Hayashi1985}, after settling into equilibrium, the gas and solid components in the disk are separated and dust grains begin sinking to the midplane to form a thin dust layer which eventually fragments into planetesimals, ingredients of planets.
The radial profile of the surface density structure at
this stage is likely one of the most important parameters to control what
kinds of planets form in the disk \citep{bib:Kokubo2002}.\par

Since the gas and dust are well mixed in the beginning, the grain growth and sedimentation are key processes to separate dust from gas.
Theoretical studies have suggested that dust grains are removed from the flaring surface and will sink to the midplane with growing by coalescence along the disk evolution \citep{bib:Dullemond2004,bib:Tanaka2005}.
One way to find such evolved disks is to carefully examine the Spectral Energy Distribution (SED) of young stars as well as detection of (sub)millimeter continuum emission, because the shape of the SED of young stars is sensitive to its temperature structure mainly determined by the vertical distribution of dust grains \citep{bib:Chiang1997}.
\citet{bib:Tanaka2005} studied the dust growth and settling in passive disks to investigate the evolution of disk structure and SEDs.
Their numerical results indicated that the flux densities at mid-infrared (MIR) from a disk get lower as a result of dust grain growth and settling from the surface layer.\par

In the last two decades, extensive observations from optical to millimeter wavelengths have been conducted to study disk properties ultimately determining the nature of forming planets.
Pre-main sequence stars, particularly in the quiescent phase such as weak line T Tauri stars (WTTSs) or \mbox{Class\ III} stars are good candidates for searching such evolved dusty disks in the process of sedimentation.
A pioneer survey in (sub)millimeter continuum emission of the circumstellar dust around young stars was conducted by \citet{bib:Beckwith1990} and followed by many other groups \citep{bib:Andre1994,bib:Osterloh1995,bib:Nurnberger1997,bib:Kitamura2002,bib:Andrews2005,bib:Andrews2007}.
\citet{bib:Kitamura2002} made an imaging survey of single T Tauri stars in Taurus and derived the physical properties of the disks combined with the SEDs.
They found the radial expansion of the disks attributed to outward transport of angular momentum with decreasing H$\alpha$ line luminosity, a measure of disk evolution.
\citet{bib:Andrews2005} made a large survey of disk emission of more than 100 young stars in Taurus and found that there is a distinct difference in detection rates between classical T Tauri star (CTTS; 90\%) and WTTS (15$\pm$5\%).
No massive evolved disk around WTTS, however, was found.\par

In the present paper, we report the discovery of an unusual cold massive disk around the T Tauri star, \mbox{V1094\ Sco}, at the southern tip of Lupus\ 3 at a distance of 150 pc.
The source has a quiescent accretion disk presumably in the transient stage from CTTS to WTTS because its equivalent width of H$\alpha$ ranged 7--14 $\mathrm{\AA}$ \citep{bib:Krautter1997,bib:Wichmann1999}.
\mbox{V1094\ Sco} is known as a single star whose spectral type, mass, and age are K6, 0.78 $M_\sun$, and 3 Myr, respectively \citep{bib:Wichmann1997}.



\section{Observations}
Our observations have been performed with the Atacama Submillimeter Telescope Experiment (ASTE), the \mbox{10 m} submillimeter telescope operated by Nobeyama Radio Observatory (NRO), a branch of National Astronomical Observatory Japan (NAOJ), in collaboration with University of Chile, and Japanese institutes including University of Tokyo, Nagoya University, Osaka Prefecture University, Ibaraki University, and Hokkaido University.
The telescope is located  at Pampa la Bola, Chile, an altitude of 4860m, and is remotely operated from the ASTE operation facilities in Chile and Japan through the network observation system N-COSMOS3 \citep{bib:Kamazaki2005}.
We employed two receivers, AzTEC 1.1 mm camera \citep{bib:Wilson2008} and \mbox{CATS345} receiver \citep{bib:Inoue2008}, for millimeter continuum observations and molecular line observations at submillimeter wavelengths, respectively.

\subsection{Millimeter Continuum Observations with AzTEC}
AzTEC is a 144-element bolometer array camera installed on the ASTE between April 2007 and December 2008 \citep{bib:Wilson2008}.
In this period, the array was configured to operate at a wavelength of 1.1 mm.
The angular resolution and field of view of AzTEC on ASTE were 28\arcsec and 7\farcm8, respectively.\par

We have made wide-field (40\arcmin$\times$40\arcmin) millimeter continuum observations of the \mbox{Lupus\ 3} molecular cloud including the target, \mbox{V1094\ Sco}, in July 2007.
The individual on-the-fly maps were obtained in the raster-scan mode with a scanning velocity of 100$\arcsec$ s$^{-1}$.
The telescope was scanned along the AZ-EL coordinates with a spacing of 1\farcm95, a quarter of the field of view.
All the data were taken at night or early morning and the average 270 GHz radiometric opacity at zenith was $\sim0.05$ during the observation period.
Telescope pointing was checked by mapping the quasar \mbox{J1517-243} every one and half, or two hours.
The pointing corrections are applied during the data analysis.
The flux conversion factor from optimal loading to source flux density of the AzTEC array was calibrated by observing Uranus, Neptune, or 3C279 every night.\par

The AzTEC data set was reduced by using the AzTEC Data Reduction Pipeline \citep{bib:Scott2008} for the identification of point sources.
Following \citet{bib:Scott2008}, we employed the Principle Component Analysis (PCA) cleaning method to remove the atmospheric emission.
The pixel size of the final map was set to be 3$\arcsec$, 10 \% of AzTEC beam size.
The noise level and effective resolution of the final map are 5.0 mJy beam$^{-1}$ and 35$\arcsec$, respectively.

\subsection{Sub-Millimeter Molecular Line Observations with CATS345}
$^{12}$CO and $^{13}$CO J=3--2 line observations were performed in March 2008 with a 2 sideband-separating (2SB) receiver, CATS345, on ASTE.
The half power beam widths (HPBWs) at 345.79 and 330.59 GHz were $22\arcsec$ and $23\arcsec$, respectively, which corresponds to 3300 and 3450 AU at the distance to the target.
The main beam efficiency was about 60 \% around these frequencies.
The system noise temperatures were typically 400 and 350 K for $^{12}$CO and $^{13}$CO, respectively.
For the backend, we used a 1024 channel digital auto-correlator, which has a band width of 128 MHz and a resolution of 125 kHz, corresponding to 111 and 0.11 km s$^{-1}$, respectively, at 345 GHz.
The telescope pointing calibration was performed every 1.5-2 hours by observing \mbox{CRL\ 4211}, and the resulting pointing accuracy was typically $3\arcsec$.
For the absolute flux calibration, the integrated intensities of \mbox{M17\ SW} and \mbox{IRC\ 10216} were measured two or three times per day and compared with those by \citet{bib:Wang1994}.
Data reduction and analysis were made with the NEWSTAR software package developed in NRO, which is a frontend of the Astronomical Image Processing System (AIPS) developed in the National Radio Astronomy Observatory (NRAO).\par

We employed the five-point observation method in which spectra are obtained not only at a stellar position but also at four adjacent points $22\arcsec$ apart from the stellar position in the R.A. or Dec. direction. 
This method enables us to deduce the excess emission toward the central star by subtracting the average spectrum of the surrounding four spectra from the central spectrum, even if the star is located in extended components, such as an ambient cloud.

\section{Results of the Continuum Observations}\label{sec:cont_results}
\begin{figure*}
\epsscale{1.0}
\plotone{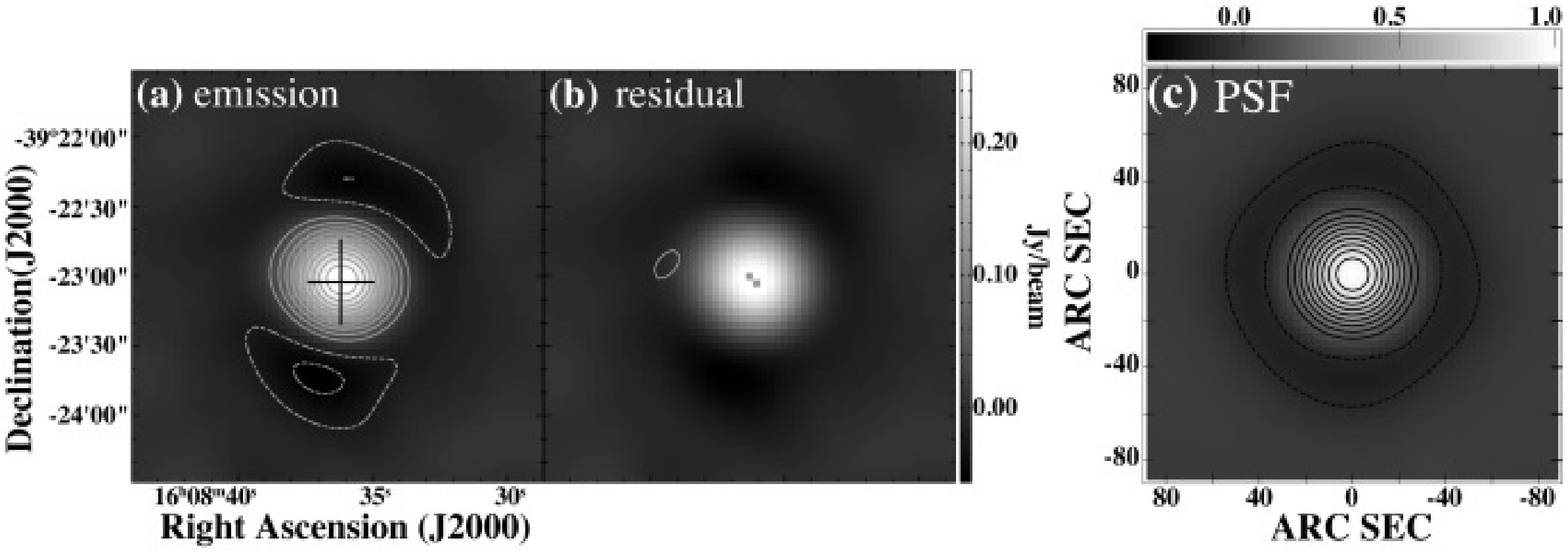}
\caption{(a)1.1 mm continuum map of \mbox{V1094\ Sco} obtained with AzTEC on
 ASTE. The contour lines start at $\pm$5$\sigma$ with intervals of 5$\sigma$
 where 1$\sigma$ is 5 mJy beam$^{-1}$. The negative levels are indicated by
 the dashed lines.
 The position of \mbox{V1094\ Sco} is shown by the central cross.
 (b) The residual contour map by deconvolving the PSF via the clean algorithm is
 superposed on the continuum map (a) in grey scale.
 The contour levels and the grey scale are the same as in Figure\
 \ref{fig:cont}(a).
 The shaded area indicates the positions of the clean components.
 (c) Point Spread Function (PSF) of the map (a) obtained with AzTEC on ASTE.
 The contour levels are $\pm$10$\times\mathrm{n}$ \% of the peak value
 (n=1,2,3,$\dots$).
 The negative levels are indicated by the dashed lines.}\label{fig:cont}
\end{figure*}

The 1.1 mm continuum map around \mbox{V1094\ Sco} is shown in Figure\ \ref{fig:cont}a.
Strong compact continuum emission was detected towards the stellar position.
From the 2-d Gaussian fitting, the peak position of the emission is obtained to be 16$^\mathrm{h}$08$^\mathrm{m}$36\fs1, $-$39\arcdeg23\arcmin1\farcs6 in J2000, which agrees well with the 2MASS position of \mbox{V1094\ Sco} within 1$\farcs$5.
The total flux density at 1.1 mm was measured to be 272.1$\pm$7.6 mJy from the deconvolved cleaned map (see below), which is one of the strongest 1.1 mm emission among T Tauri stars in the Lupus molecular cloud.
The negative contours around the emission are due to a point spread function (i.e., a beam; see Figure\ \ref{fig:cont}c) with a negative hole generated from the PCA cleaning method.\par

Figure\ \ref{fig:cont}b shows a deconvolved map by subtracting the measured point spread function (PSF) from the emission via the \textit{clean} algorithm.
The PSF of the AzTEC map for \mbox{V1094\ Sco} is shown in Figure\ \ref{fig:cont}c.
From the deconvolved map, we found that the emission is composed largely of a spatially unresolved component.
The derived clean components are located around the star over two pixels ($3\arcsec \times \sqrt{2} \sim 4\farcs2$ in size) with its separation comparable to the pointing uncertainty estimated from the pointing measurements (Figure\ \ref{fig:cont}b).
Therefore, we consider that the strong millimeter emission comes from a compact ($r\le 4\farcs2/2 =320$ AU) circumstellar dusty disk associated with \mbox{V1094\ Sco}.
This is supported by the fact that, unlike the case of a stellar type flare seen in WTTS \citep{bib:Gudel2002}, \mbox{V1094\ Sco} did not show any time variation of the flux density at millimeter wavelengths during the observing period.\par

Figure\ \ref{fig:sed} shows the SED of \mbox{V1094\ Sco}.
The flux density at 1.1 mm is unusually strong as the 1.1 mm to infrared flux density ratio is near unity and larger than those for T Tauri stars in Taurus \citep{bib:Strom1989,bib:Beckwith1990}.
Adapting the extinction relation of \citet{bib:Rieke1985}, we measured the spectral index for the dereddened fluxes in the range from 2 to 10 $\micron$, $\alpha_{2,10}$, to be $-1.39\pm0.10$, where $\alpha =[d\log(\lambda F_\lambda)/d\log(\lambda)]$.
This value suggests that \mbox{V1094\ Sco} should be categorized as a \mbox{Class\ II} object.
The shape of the measured SED is unique in the sense that there are not distinct broadband spectral features, such as a large dip, from near-infrared (NIR) to millimeter wavelengths with a power-law like monotonic decrease.
The index of the decrease is measured to be $-0.98\pm0.02$ in the range from 2 to 1100 $\micron$.
However, there is a small turnover around 10 $\micron$, i.e., the index at the longer wavelength, from 20 to 1100 $\micron$, follows nearly the entire trend to be $-0.93\pm0.03$ while that from 2 to 10 $\micron$ is $-1.39\pm0.10$.
Furthermore, there is no dip feature in the NIR to MIR wavelength range caused by clearing material in the inner parts of the disk, such as \mbox{GM\ Aur} \citep{bib:Hughes2009}.\par


\section{Results of the Molecular Line Observations}
\begin{figure*}
\epsscale{1.0}
\plotone{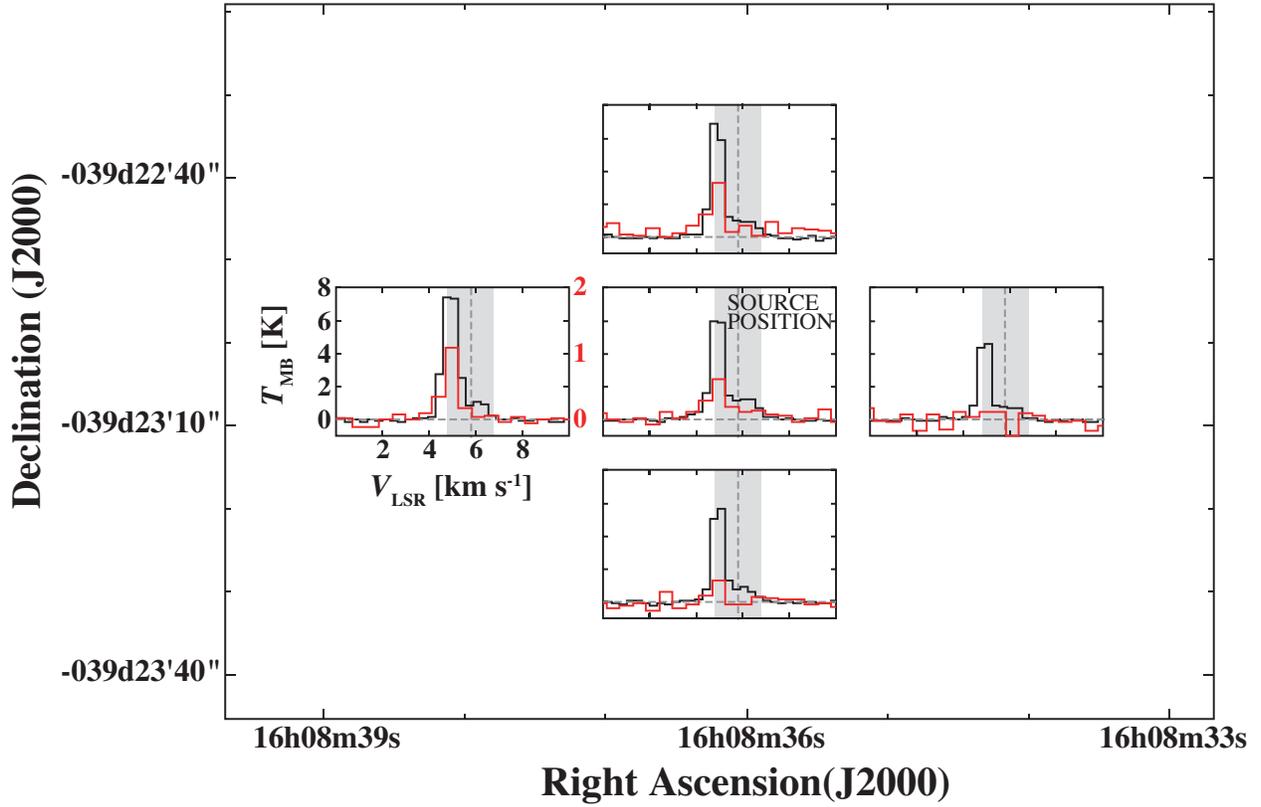}
\caption{$^{12}$CO (black) and $^{13}$CO (red) J=3--2 five-point profile
 maps of \mbox{V1094\ Sco}. The observed points are located at the
 stellar position, to the 22$\arcsec$ north, south, east, and west from
 the center. The spectra are smoothed in resolutions of 0.3 and 0.5 km
 s$^{-1}$ for $^{12}$CO and $^{13}$CO, respectively. The shaded areas
 show the radial velocity of the central star with their error
 \citep{bib:Guenther2007}.}\label{fig:5point}
\end{figure*}

The five-point spectra of the $^{12}$CO and $^{13}$CO J=3--2 emissions are shown in Figure\ \ref{fig:5point}.
The significant $^{12}$CO and $^{13}$CO emissions were detected over the five-points around the star.
There is a clear intensity gradient along the east--west direction in the $^{13}$CO spectra as well as in $^{12}$CO.
In the $^{12}$CO spectra, we could not find any symmetrical wing feature of an outflow with a velocity width of $\sim$10 km s$^{-1}$ which is often seen around a low-mass young stellar object.
The $^{12}$CO and $^{13}$CO spectra are peaked around the LSR velocity of 4.9 km s$^{-1}$, which is slightly shifted from the systemic velocity of the \mbox{Lupus\ III} main cloud \citep[4.1 km s$^{-1}$;][]{bib:Tachihara1996}.
The peak intensity in $T_\mathrm{MB}$ scale and the velocity width in FWHM are 7.0 K and 0.66 km s$^{-1}$, respectively, on average over the five-points for $^{12}$CO, and 0.7 K and 0.85 km s$^{-1}$ for $^{13}$CO.\par

From the intensity ratio of the $^{12}$CO to $^{13}$CO spectra, we estimate the optical depth of both lines toward the star.
If we assume the same excitation temperature, the mean optical depths of $^{12}$CO and $^{13}$CO are obtained to be 16.7 and 0.19, respectively.
Since the $^{13}$CO emission is optically thin in contrast to $^{12}$CO, the associated gas with the star can be expected to be seen in $^{13}$CO through the parent cloud as excess emission towards the star.\par

By assuming the local thermodynamic equilibrium with the excitation temperature, $T_\mathrm{ex}$, the H$_2$ mass of the gas in the beam solid angle toward the source, $M_\mathrm{H_2}$, can be estimated from the intensity of the optically thin $^{13}$CO J=3--2 line as follows:
\begin{eqnarray}
\nonumber M_\mathrm{H_2} &=& 1.99\times10^{-14}
\frac{1}{X_\mathrm{^{13}CO}} \biggl( \frac{D}{[\mathrm{pc}]}
\biggr)^2\\
\nonumber &\times&
\frac{\exp(15.9/T_\mathrm{ex})}{1-\exp(-15.9/T_\mathrm{ex})}
\frac{T_\mathrm{ex}}{J(T_\mathrm{ex})-0.05} \\
&\times&  \frac{\tau_\mathrm{^{13}CO}}{1-\exp(-\tau_\mathrm{^{13}CO})}
\int \frac{T_\mathrm{A}^\ast(\mathrm{^{13}CO})}{\eta_\mathrm{MB}} dv  \
\ \ \ \ [M_\sun]\mathrm{,} \label{eq:mass}
\end{eqnarray}
where 
$
J(T_\mathrm{ex})=T_0/[\exp(T_0/T_\mathrm{ex})-1]
$,
$
T_0 = h \nu /k_\mathrm{B}
$,
$X_\mathrm{^{13}CO}$ is the fractional abundance of $^{13}$CO with respect to H$_2$, $D$ is the distance to the source, $\tau$ is the optical depth, and $\eta_\mathrm{MB}$ is the main beam efficiency of ASTE \citep{bib:Scoville1986}.
We adopt an excitation temperature 10 K which is the typical value in the ambient molecular cloud \citep{bib:vilas-boas2000}, and $1.7\times10^{-6}$ for $X_\mathrm{^{13}CO}$ by assuming that the abundance ratio of $\mathrm{^{12}CO/^{13}CO}$ and the $\mathrm{H_2}/\mathrm{^{12}CO}$ conversion factor are 60 and $10^4$, respectively.
Our mass as well as column density are estimated to be $6.7\times10^{-3}$ $M_\sun$ and $1.2\times10^{21}$ cm$^{-2}$ by using equation\ (\ref{eq:mass}).
The column density is much smaller than the averaged one in the Lupus\ 3 cloud \citep{bib:Tachihara1996}, suggesting that there is no large and massive envelope around \mbox{V1094\ Sco}.\par

\begin{figure}
\epsscale{1.0}
\plottwo{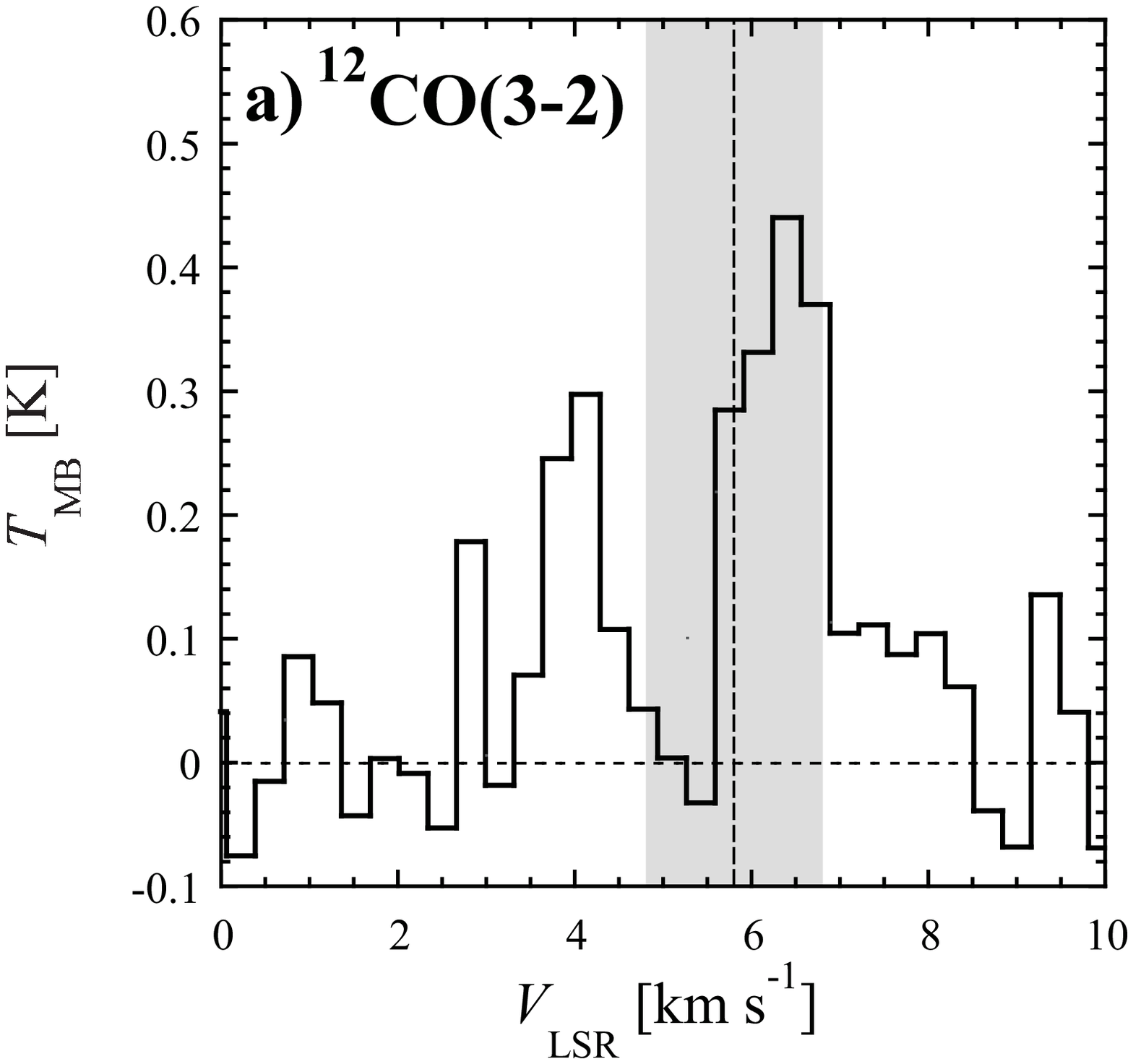}{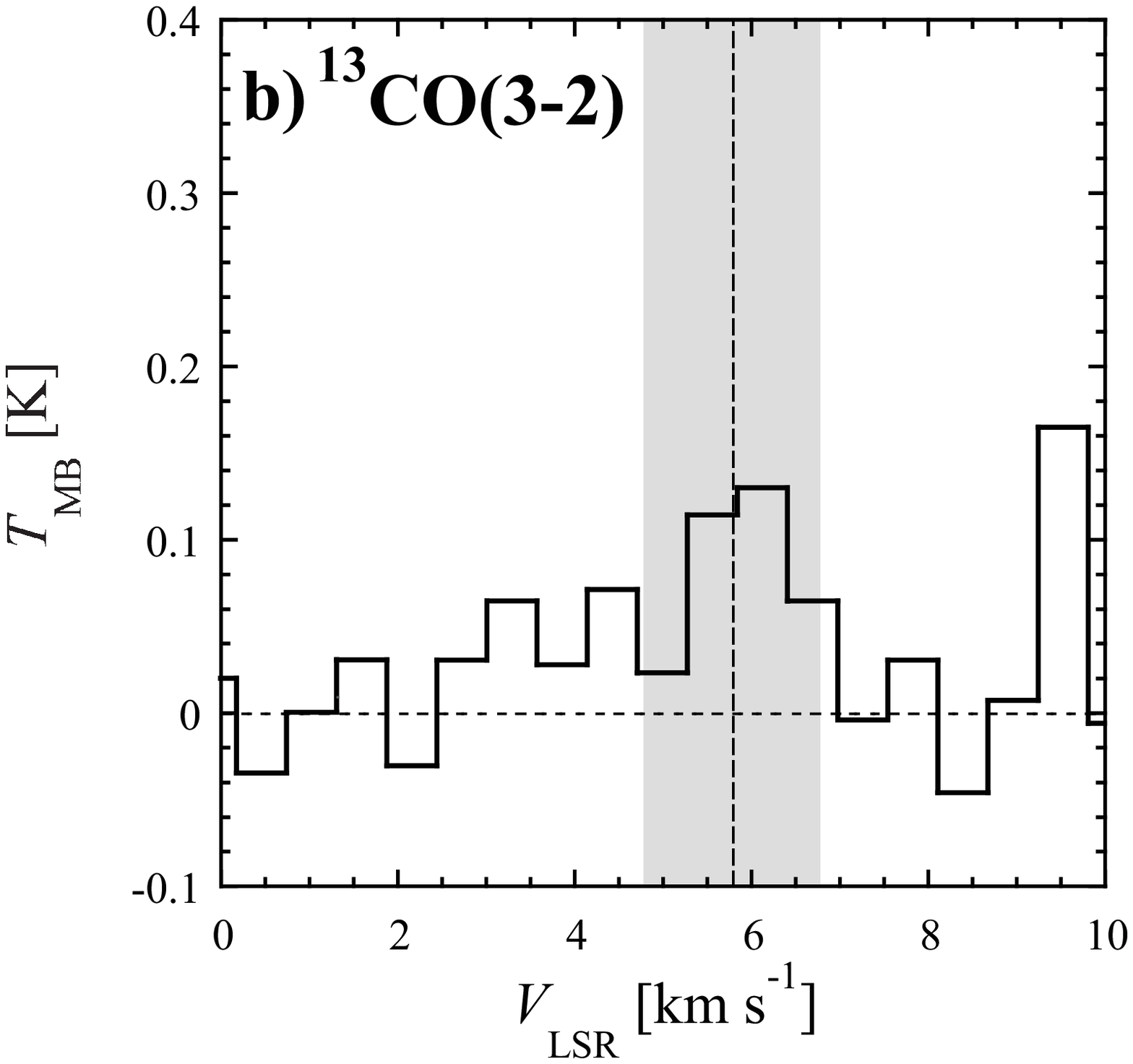}
\caption{
 Residual spectra of $^{12}$CO and $^{13}$CO towards \mbox{V1094\ Sco}.
 The vertical dotted lines and the shaded areas show the radial
 velocity of the star and its uncertainty
 \citep{bib:Guenther2007}.}\label{fig:residual}
\end{figure}

To search for the compact excess emission towards the star, we made residual spectra of the $^{12}$CO and $^{13}$CO emission as shown in Figure\ \ref{fig:residual}.
Each residual spectrum was obtained by subtracting the average of the surrounding four spectra from the central one.
The excess emission above the 2$\sigma$ noise level over two velocity channels can be discerned in the $^{13}$CO residual spectrum around the velocity of 6 km s$^{-1}$.
The peak intensity and the velocity width in FWHM are measured to be 0.12 K and 1.43 km s$^{-1}$, respectively.
Since the velocity of the excess emission agrees well with the radial velocity of the star, 5.8$\pm$1.0 km s$^{-1}$ \citep{bib:Guenther2007}, this may be the possible detection of the gas disk around \mbox{V1094\ Sco}.\par

The $^{12}$CO residual emission was detected at $V_\mathrm{LSR}=$ 4.1 and 6.4 km s$^{-1}$ whose peak intensities are 0.30 and 0.43 K in $T_\mathrm{MB}$, respectively.
Although the $^{12}$CO emission is shown to be optically thick, the double-peaked profile of the residual spectrum towards the star might suggest the presence of a rotating gas disk around \mbox{V1094\ Sco}.
We tried to fit the rotating disk model \citep{bib:Kitamura1993} to the both spectra to examine quantitatively.
We could not, however, find any parameter set to reproduce the both residual spectra consistently. 
For example the $^{12}$CO intensity can be reproduced for the larger rotating disk model.
Such a disk model expects the stronger $^{13}$CO intensity than the obtained and can not explain the both intensities compatibly.
In addition, both velocity widths are not well fitted by the model simultaneously.
Consequently, the residual spectrum in $^{12}$CO is not likely to originate from the circumstellar gas disk.

\section{Discussion}

\subsection{Analysis of Spectral Energy Distribution}\label{sec:sed_fit}
The measured SED of \mbox{V1094\ Sco} has the unique feature: there is no significant turnover and a monotonic decrease from NIR to millimeter.
This feature is well described by a flattened disk for the dust component.
According to \citet{bib:Chiang1997}, the flattened disk has a lower temperature than that of a flared disk with the well-mixed dust because the central star can not illuminate the flattened disk efficiently.
In this case, the SED is predicted to have an asymptotic power-law slope with an index of $n=4/3$ from 30 $\micron$ to 1 mm.
The MIR to millimeter continuum radiation mainly comes from the optically thick interior of the disk.
On the other hand, the SED of a flared disk rises in the MIR to far-infrared (FIR) regime owing to the strong emission from a hot surface layer and gradually falls down toward the millimeter regime, mainly tracing the cooler interior.\par

To deduce disk parameters, we performed the least square fitting to the SED of \mbox{V1094\ Sco} using a simple power-law disk model \citep[same as model\ 1 case in][]{bib:Kitamura2002}.
Our disk model is assumed to have a power-law form of the surface density and temperature variations over the entire area of the disk, $\Sigma(r)$ and $T(r)$, respectively: 
\begin{equation}
 \Sigma(r) = \Sigma_\mathrm{100AU} \Biggl(\frac{r}{\mathrm{[100AU]}} \Biggr)^{-p}
\end{equation}
and
\begin{equation}
 T(r) = T_\mathrm{1AU} \Biggl(\frac{r}{\mathrm{[1AU]}}\Biggr)^{-q} \mathrm{.}
\end{equation}
Here, $r$ is the radial distance in units of AU, $\Sigma_\mathrm{100AU}$ is the surface density at 100 AU, and $T_\mathrm{1AU}$ is the temperature at 1 AU.
We applied the model to the data without the extinction correction because it was corrected in the model by calculating the mass opacity at each frequency from $A_\mathrm{v}$ with the extinction curve \citep[figure\ 1 of][]{bib:Adams1988}.
The lower limit of the temperature is fixed at 10 K which is the typical temperature of the Lupus\ 3 cloud \citep{bib:vilas-boas2000}.
We adopt the effective temperature of the star of 4255 K, the stellar mass of 0.78 $M_\sun$ and the visual extinction toward the star of 1.75 mag \citep{bib:Wichmann1997}.
The outer radius of the disk, $R_\mathrm{out}$, and the disk inclination angle, $i$, are fixed to be 320 AU and 45 degree, respectively.
The outer radius is derived from the pixel size of 3$\arcsec$ as an upper limit because the emission was not spatially resolved.
Stellar radius, $R_\ast$, inner radius of the disk, $R_\mathrm{in}$, $T_\mathrm{1AU}$, $q$ and $\Sigma_\mathrm{100AU}$ are the free parameters of the disk model.
The model SEDs are computed and fitted to the observational data for the following four cases: $p=1$ and $1.5$ \citep[minimum mass solar nebula by][]{bib:Hayashi1985} and $\beta=0$ and 1, where $\beta$ is the power-law index of the dust mass opacity coefficient, $\kappa_\nu = 0.1\times(\frac{\nu}{10^{12}\mathrm{Hz}})^\beta$ cm$^2$ g$^{-1}$ \citep{bib:Beckwith1990}.
In the fitting, we treated all the data points with the equal weights of the largest uncertainty because the SED fitting tends to be biased toward the data points with small uncertainties at shorter wavelengths.\par

\begin{figure}
\epsscale{0.8}
\plotone{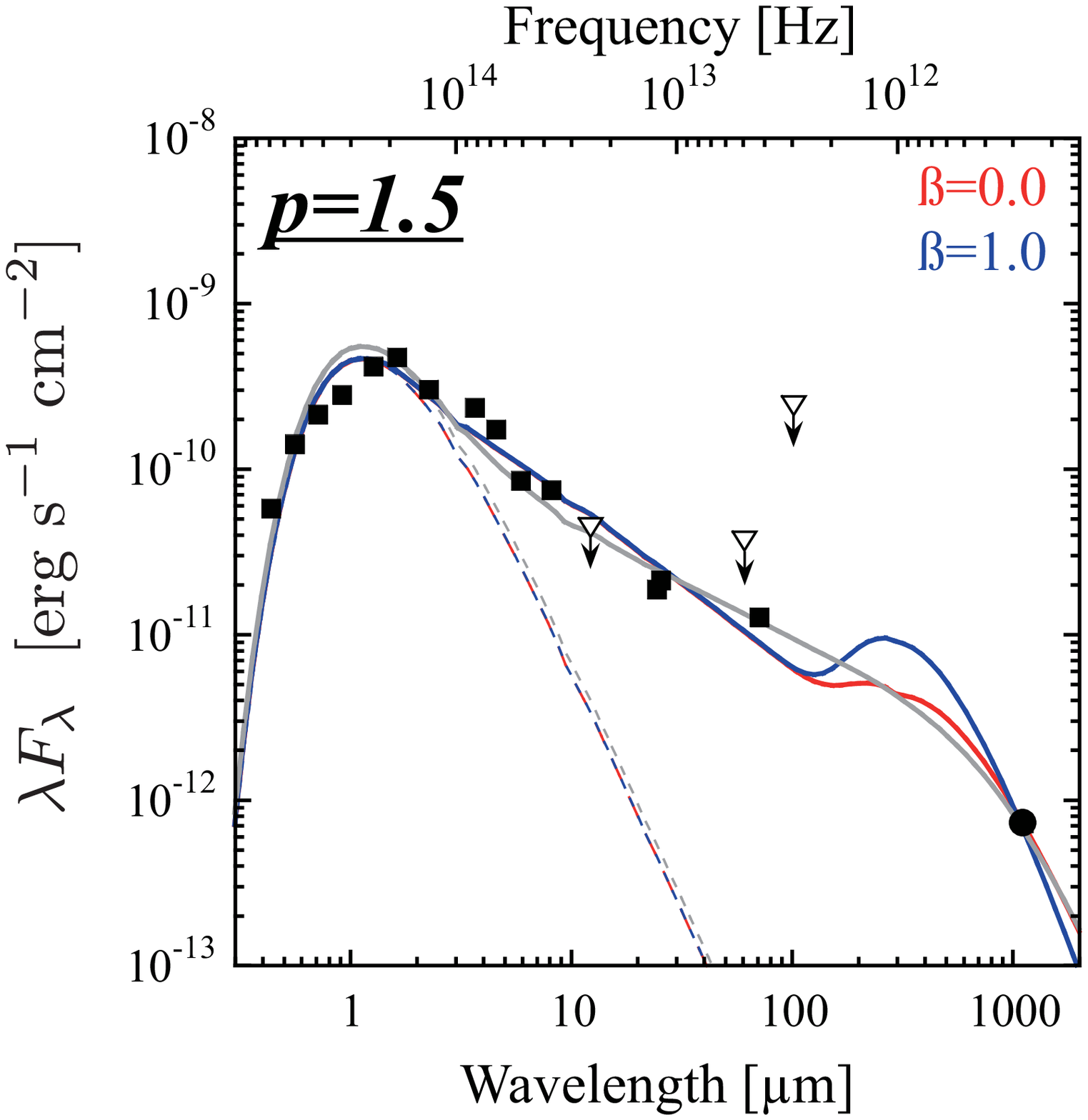}
\caption{Spectral Energy Distribution (SED) of \mbox{V1094\ Sco}. The
 result of our observations is shown in the filled circle. The filled
 squares indicate the SED compiled from previous studies; the NOMAD catalog
 \citep{bib:Zacharias2005}, the DENIS database
 \citep{bib:DenisConsortium2005}, the 2MASS point source catalog
 \citep{bib:Cutri2003}, the {\it IRAS} faint source catalog
 \citep{bib:Moshir1992}, and the spitzer photometry \citep[3.6, 4.5,
 5.8, 8.0, and 24 $\micron$][]{bib:Cieza2007}.
 The flux density at 70 $\micron$ with {\it Spitzer} MIPS was measured
 by analyzing the  archived data using the program MOPEX/APEX software packages.
 The {\it IRAS} flux densities at 12, 60, and 100 $\micron$ are
 upper limits and shown by the open triangles.
 The dashed lines show the stellar contributions and the solid line show
 the best-fitted curves in the case of $p=1.5$.
 The red and blue lines show the calculations for $\beta=$0 and 1,
 respectively.
 A grey line indicates the fitting results for the model disk with a low
 threshold temperature of 2.7 K in the temperature profile when $p=1.5$ and
 $\beta=0$.
 }\label{fig:sed}
\end{figure}

These model SEDs seem to agree with the observed SED as shown in Figure\ \ref{fig:sed} and its reduced chi-squares are measured to be $\sim9$.
The derived parameters by the fitting are listed in Table\ \ref{tab:fitting} and \ref{tab:fitting_result}.
In Figure\ \ref{fig:sed}, we plotted the fitting results only for $p=1.5$ case because the two best-fit curves with $p=1$ and 1.5 are almost identical for each $\beta$ value.
Two remarkable disk features are derived in the fitting.
First, the disk is very cold with a temperature of 99 K at 1 AU.
The fitted stellar luminosity of 0.85 $L_\sun$ suggests that $T_\mathrm{1AU}$ in the condition of radiative equilibrium is expected to be 269 K when the disk is in hydrostatic equilibrium and gas and dust are well mixed.
Previous studies of the SED fitting have found that most T Tauri stars in Taurus have a higher temperature of 100--400 K at 1 AU \citep{bib:Kitamura2002,bib:Andrews2005}.
Our derived low temperature suggests a flattened dust disk as predicted by theoretical studies \citep[e.g.][]{bib:Chiang1997}.
The second remarkable finding is that the disk is massive.
Our estimated disk mass ranges from 0.03 to 0.12 $M_\sun$ and the ratio of the disk to stellar mass is 3--13\%.
These values are much higher than the typical values for T Tauri stars in the Taurus-Auriga region: median disk masses are $4\times10^{-3}$ and $2\times10^{-3}$ $M_\sun$ for CTTS and WTTS, respectively, and the median disk-to-star mass ratio is only 0.5 \% \citep{bib:Andrews2005}.
It is very rare to find a massive disk in a quiescent accretion phase like a WTTS one.
For example, \citet{bib:Andrews2005} have found that only 15 \% of WTTSs are detectable at the submillimeter wavelengths.\par

The above arguments are not affected much by either the disk inclination angle or the disk outer radius.
If we change the disk inclination angle from 20 to 70 degrees with $R_\mathrm{out}=320$ AU and $p=1.5$, the temperature at 1 AU varies from 90 to 128 K, which is still low, and the index $q$ remains around 0.68.
The variation of the disk mass is less than 20 \%, remaining the massive disk.
Furthermore, when we adopt a smaller $R_\mathrm{out}$ than 320 AU, a higher surface density is required to reproduce the observed SED, resulting the massive disk.
For example, in the $R_\mathrm{out}$=110 AU case, $\Sigma_\mathrm{100AU}$ becomes more than 5 times larger than that for the $R_\mathrm{out}$=320 AU case and the disk mass increases up to 0.42 $M_\sun$, which remains massive.
For a smaller opaque disk, we cannot accurately estimate the disk mass, though our interpretation of the massive disk is still valid.
Therefore, we definitely conclude that the disk around \mbox{V1094\ Sco} is colder than the usual flared disk and that the disk mass is much larger than the typical value of T Tauri stars in Taurus.\par


In our SED fitting, we adopted a low threshold of 10 K in the temperature profile because the source is likely embedded in its parent cloud as shown by our $^{12}$CO observations.
This threshold causes an increment of flux densities around at 300 $\micron$ in the model SED because the temperature of 10 K corresponds to 290 $\micron$ by Wien's law (see Figure\ \ref{fig:sed}).
To investigate the effect of the temperature threshold, we also analyzed the SED by using a disk model with a threshold of 2.7 K, the temperature for the cosmic background radiation.
In this case, the observed SED can be well fitted only when the disk inclination angle is less than 30 degree.
The fitted SED for $p=1.5$ and $\beta=0.0$ case is shown by a gray line in Figure\ \ref{fig:sed}; the best-fit curve can also explain the observed SED, and the disk mass becomes four times larger than that in the 10 K threshold case.
Note that such a massive disk relative to the stellar mass is likely to be gravitationally unstable.
On the other hand, $T_\mathrm{1AU}$ and $q$ decrease only by 2 \% and 11 \%, respectively.
Consequently, the threshold in the temperature profile of the disk model does not change our interpretation of a cold massive disk around \mbox{V1094\ Sco}.\par

\subsection{Rare Object to Provide the Initial Conditions of Planet Formation}

The cold and massive disk we found around \mbox{V1094\ Sco} strongly suggests the presence of a flattened disk where a large number of small dust grains are settling.
Since \mbox{V1094\ Sco} seems to be in the quiescent phase of disk accretion, the majority of dust particles are expected to just settle into the midplane with dust growth through coalescence. 
Theoretically, the thin dust layer in the midplane fragments into numerous planetesimals by its gravitational instability, and as a result, a hole appears at the inner part of the disk to produce a dip feature at $\sim10$ $\micron$ in SED. 
However, we could not find such a feature at the IR wavelengths in the SED of \mbox{V1094\ Sco}, indicating that there is no inner hole created by the formation of planetesimals. 
Indeed, the strong continuum emission toward \mbox{V1094\ Sco} that we detected suggests that a large number of small particles rather than larger planetesimals with sizes of $\sim10$ km is the dominant component in the circumstellar disk. 
Consequently, the disk around \mbox{V1094\ Sco} is most likely in the quiescent stage after the dust sedimentation in the midplane, but just prior to the planetesimal formation.\par

\mbox{V1094\ Sco} is a rare object in terms of the shape of its SED.
To find a similar object with a massive cold disk to \mbox{V1094\ Sco}, we checked the available photometric data at the MIR and millimeter wavelengths \citep{bib:Strom1989,bib:Beckwith1990} and measured the flux ratio of MIR flux to millimeter flux.
We found that only \mbox{CY\ Tau}, a Class II CTTS in Taurus, has a high intensity ratio of the 1 mm to 70 $\micron$ as well as to 25 $\micron$ ($>0.5$), suggesting the strong millimeter flux despite its weak MIR flux and the SED with a monotonic slope from MIR to millimeter.
The temperature at 1 AU and the disk mass of \mbox{CY\ Tau} are estimated to be 87 K and 0.13 $M_\sun$, respectively \citep{bib:Kitamura2002}.
We can roughly estimate the timescale for the duration of a cold and massive disk like those around \mbox{V1094\ Sco} and \mbox{CY\ Tau} as follows.
Given about 30 samples in Taurus whose flux densities at 1 mm and MIR were previously determined, the timescale of the cold massive disk is estimated to be at least one order of magnitude shorter than the typical age of a T Tauri star ($\sim 1 \times10^{6-7}$ yr), if we assume that new stars randomly form with a constant rate.
Such a short timescale is consistent with the gravitational instability of the thin dust layer after the dust sedimentation at the midplane.
It is suggested that the dust layer produced by settling to the equatorial plane becomes gravitationally unstable and rapidly fragments into numerous planetesimal \citep[$\ll$10$^5$ yrs;][]{bib:Dominik2007}.\par


It is very interesting that the physical properties of such a quiescent disk provide the initial conditions of the successive formation of planets \citep[e.g.][]{bib:Kokubo2002}. 
In particular, it is crucial to reveal the radial variation of the disk surface density with high-resolution and high-sensitive observations in future. 
For the dust coagulation, it is theoretically expected that the index of the dust opacity coefficient, $\beta$, is lower than unity as a result of dust growth \citep{bib:Miyake1993}.
We treated $\beta$ as fixed parameter in the SED fitting because the photometric data at the millimeter and sub-millimeter wavelengths are limited. 
More (sub-)millimeter data are required to determine the $\beta$ value accurately.
Additionally, high-resolution molecular line observations will reveal the disk kinematics and the initial conditions of the planet forming disk more quantitatively.\par

\section{Summary}
We have detected the strong 1.1 mm continuum emission toward the T Tauri star, \mbox{V1094\ Sco}, in the Lupus molecular cloud with AzTEC on ASTE.
In addition, we have obtained the $^{12}$CO and $^{13}$CO J=3--2 spectra at the five positions around the source.
The main results of our observations are summarized in the following.

\begin{itemize}
        
 \item{Significant continuum emission was detected just toward the
      stellar position. The emission is not spatially resolved with the
      effective ASTE beam size of 35$\arcsec$, indicating that a dust
      disk is associated with the star. The total flux density is 272 mJy at
      1.1 mm which is the strongest among T Tauri stars in the Lupus molecular cloud.}
      
 \item{The $^{12}$CO and $^{13}$CO emissions were detected over the
      five-points around the star. The five-points spectra show no
      signature of a large-scale molecular outflow or a massive envelope.
      We made residual spectra by subtracting the average
      spectrum of the surrounding four spectra from the central
      spectra. The $^{12}$CO excess emission was found and the
      $^{13}$CO one was marginally discerned towards the star. The
      $^{13}$CO excess may come from the gas disk while the $^{12}$CO
      excess emission detected can not be explained by the gas disk.}
 \item{The observed SED shows two prominent features. First, there is no
      significant turnover with a constant slope from NIR to
      millimeter which is well reproduced by a flattened
      disk. Second, there is no signature of a dip at around 10 $\micron$
      suggesting no inner hole in the disk.}
 \item{We fit a power-law disk model to the SED when $p$ is fixed to be 1.0
      and 1.5, and $\beta$ is 0.0 and 1.0. The best fit parameters
      of the disk are $\Sigma_\mathrm{100AU}=$(1.25--4.97) g cm$^{-2}$,
      $T_\mathrm{1AU}=$99 K, $q=0.68$, and
      $M_\mathrm{disk}=$(0.03--0.12) $M_\sun$. The disk temperature at 1
      AU is lower than that of a flared disk in hydrostatic equilibrium.
      The disk mass is much higher than the typical values for T Tauri stars.}
      
 \item{The cold and massive disk around \mbox{V1094\ Sco} can be
      well interpreted as a partially settled disk of a large number of
      small dust grains.
      Together with no signature of an inner hole in the SED,
      it is most likely that the disk is in the quiescent phase after
      the dust sedimentation in the midplane, but just prior to the
      planetesimal formation. The \mbox{V1094\ Sco} may be a rare
      sample to investigate the initial conditions for the formation of
      a planetary system.}
      
\end{itemize}

\acknowledgements
We appreciate the referee for the constructive comments that have helped to improve this manuscript.
We acknowledge the ASTE and AzTEC staff for the operation and maintenance of the observing instruments.
We are grateful to \mbox{S.\ Komugi}, \mbox{T.\ Minamidani}, and \mbox{K.\ Fukue} for testing the data reduction software.
We thank \mbox{N.\ Ukita}, \mbox{B.\ Hatsukade}, and \mbox{S.\ Ikarashi} for the pointing model correction of AzTEC on ASTE.
Observations with ASTE were in part carried out remotely from Japan by using NTT's GEMnet2 and its partner R\&E (Research and Education) networks, which are based on AccessNova collaboration of University of Chile, NTT Laboratories, and NAOJ.
This work is supported in part by a Grant-in-Aid for Scientific Research (A) from the Ministry of Education, Culture, Sports, Science and Technology of Japan (No.\ 19204020) and by JSPS (No.\ 18204017).

\newpage

\begin{deluxetable}{cccccccccccccccc}
\tablecaption{Adopted parameters for SED fitting}
\tablewidth{0pt}
\tablehead{
\colhead{parameter} & \colhead{value}
}
\startdata
$T_\mathrm{eff}$ [K]                  &       4255       \\
$M_\ast$ [$M_\sun$]                   &       0.78       \\
$A_\mathrm{V}$ [mag]                  &       1.75       \\
$R_\mathrm{out}$ [AU]                 &       320        \\
$i$ [degree]                          &       45         \\
$p$                                   &       1.0 \& 1.5 \\
$\beta$                               &       0.0 \& 1.0 \\
\enddata
\label{tab:fitting}
\end{deluxetable}

\begin{deluxetable}{lcccc}
\tablecaption{Results of SED fitting}
\tablewidth{0pt}
\tablehead{
\colhead{parameter} & \multicolumn{4}{c}{value}
}
\startdata
$\beta$ & \multicolumn{2}{c}{0.0} & \multicolumn{2}{c}{1.0} \\
$p$ & 1.0 & 1.5 & 1.0 & 1.5 \\
\tableline
$R_\ast$ [AU] & \multicolumn{4}{c}{0.008} \\
$L_\ast$ [$L_\sun$] & \multicolumn{4}{c}{0.85} \\
$R_\mathrm{in}$ [AU] & \multicolumn{4}{c}{0.02} \\
$\Sigma_\mathrm{100AU}$ [g cm$^{-2}$] & 1.25 & 1.47 & 4.31 & 4.97        \\
$T_\mathrm{1AU}$ [K]                  & 99.0 & 99.0 & 99.3 & 99.4        \\
$q$                  & \multicolumn{4}{c}{0.68} \\
$M_\mathrm{disk}$ [$M_\sun$]          & 0.03 & 0.04 & 0.10 & 0.12        \\
\enddata
\label{tab:fitting_result}
\end{deluxetable}


\end{document}